\documentclass[reprint,amsmath,amssymb,aps,pra,superscriptaddress]{revtex4-1}
\usepackage{amsmath,epsfig,wrapfig,graphicx,bm,color}
\usepackage[top=0.8in, bottom=0.8in, left=0.75in, right=0.75in]{geometry}
\usepackage[utf8]{inputenc}
\usepackage{hyperref}

\begin{document}
\title{Magnetic order in the chemically-substituted frustrated antiferromagnet CsCrF$_{4}$}

\author{Shohei Hayashida}
\email{shoheih@phys.ethz.ch}
\altaffiliation[Present address: ]{Laboratory for Solid State Physics, ETH Z{\"u}rich, 8093 Z{\"u}rich, Switzerland}
\affiliation{Institute for Solid State Physics, The University of Tokyo, Kashiwa, Chiba 277-8581,  Japan}
\author{Masato Hagihala}
\affiliation{Institute for Solid State Physics, The University of Tokyo, Kashiwa, Chiba 277-8581,  Japan}
\affiliation{Neutron Science Division, Institute of Materials Structure Science, High Energy Accelerator Research Organization, Tsukuba, Ibaraki 305-0801, Japan}
\author{Maxim Avdeev}
\affiliation{Australian Nuclear Science and Technology Organisation, New Illawarra Rd, Lucas Heights, NSW 2234, Australia}
\affiliation{School of Chemistry, The University of Sydney, Sydney, NSW 2006, Australia}
\author{Yoko Miura}
\affiliation{Suzuka National College of Technology, Suzuka, Mie 510-0294, Japan}
\author{Hirotaka Manaka}
\affiliation{Graduate School of Science and Engineering, Kagoshima University, Korimoto, Kagoshima 890-0065, Japan}
\author{Takatsugu Masuda}
\affiliation{Institute for Solid State Physics, The University of Tokyo, Kashiwa, Chiba 277-8581,  Japan}
\affiliation{Trans-scale Quantum Science Institute, The University of Tokyo, Tokyo 113-0033, Japan}

\date{\today}

\begin{abstract}
The effect of chemical substitution on the ground state of the geometrically frustrated antiferromagnet CsCrF$_4$ has been investigated through a neutron powder diffraction experiment. 
Magnetic Fe-substituted CsCr$_{0.94}$Fe$_{0.06}$F$_{4}$ and nonmagnetic Al-substituted CsCr$_{0.98}$Al$_{0.02}$F$_{4}$ samples are measured, and magnetic Bragg peaks are clearly observed in both samples. 
Magnetic structure analysis revealed a 120$^{\circ}$ structure having a magnetic propagation vector ${\bm k}_{\rm mag}=(0,0,1/2)$ in CsCr$_{0.94}$Fe$_{0.06}$F$_{4}$.
For CsCr$_{0.98}$Al$_{0.02}$F$_{4}$, a quasi-120$^{\circ}$ structure having ${\bm k}_{\rm mag}=(1/2,0,1/2)$ is formed.
It is notable that the identified magnetic structure in CsCr$_{0.94}$Fe$_{0.06}$F$_{4}$ belongs to a different phase of ground states from those in CsCr$_{0.98}$Al$_{0.02}$F$_{4}$ and the parent CsCrF$_{4}$.
These results suggest that the Fe substitution strongly influences the ground state of CsCrF$_{4}$.
\end{abstract}

\maketitle

\section{Introduction}
Geometrically frustrated magnets have attracted great interest in condensed matter physics because of their exotic magnetic states induced by macroscopic degeneracy of magnetic states at low temperatures~\cite{Balents2010,Savary2017}.
Since the macroscopic degeneracy in the low-energy region can be lifted even by small perturbations, geometrical frustration highlights small effects such as single-ion anisotropy~\cite{Moessner1998,Melchy2009}, 
the Dzyaloshinskii-Moriya interaction~\cite{Elhajal2002,Elhajal2005}, 
magnetic dipole-dipole interaction~\cite{Hertog2000,Maksymenko2015},
exchange randomness~\cite{Saunders2007,Andreanov2010,Watanabe2014,Kawamura2014}, 
and site dilution~\cite{Garcia2001,Maryasin2013,Andreanov2015}.
These may play key roles in determining ground states in frustrated magnets.

\begin{figure}[tbp]
\includegraphics[scale=1]{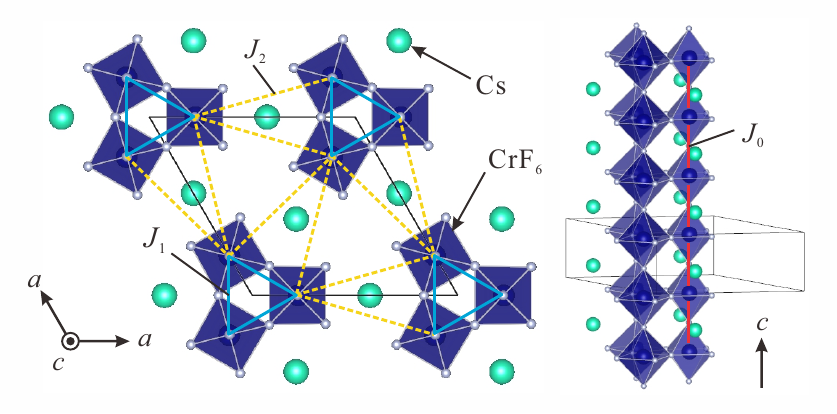}
\caption{Schematic view of the crystal structures of CsCrF$_{4}$ (hexagonal, space group $P\overline{6}2m$).
Red lines indicate the nearest-neighbor interaction along the $c$ axis, $J_{0}$.
Blue and yellow lines indicate the nearest- and second-neighbor interactions in the
$ab$ plane, $J_{1}$ and $J_{2}$.}
\label{fig1}
\end{figure}

The equilateral triangular spin tube antiferromagnet CsCrF$_{4}$ is one of the intriguing species in geometrically frustrated magnets~\cite{Manaka2009,Manaka2011_1}.
It crystallizes in a hexagonal structure with the space group $P\overline{6}2m$ as illustrated in Fig.~\ref{fig1}.
The magnetic properties are due to $S=3/2$ Cr$^{3+}$ ions.
It is unique that equilateral triangles formed by CrF$_{6}$ octahedra are stacked along the crystallographic $c$ axis, forming triangular spin tubes.
These tubes magnetically couple with one another and form the kagome-triangular lattice in the $ab$ plane~\cite{Ishikawa2014,Seki2015}.
Magnetic susceptibility measurements revealed the Curie-Weiss temperature $\theta_{\rm CW}=-145$~K and a broad maximum at $T\sim 60$ K indicative of developing short-range antiferromagnetic spin correlations~\cite{Manaka2009,Manaka2011_1}.
Due to the geometrical frustration and low dimensionality of the triangular spin tube, no clear evidence of a magnetic phase transition was found in thermodynamic and magnetic measurements~\cite{Manaka2009,Manaka2011_1,Manaka2011_2,Manaka2013,Manaka2015}. 

In a breakthrough, recent neutron powder diffraction study identified long-range magnetic order of CsCrF$_{4}$ below 2.8~K.
The magnetic moments form a quasi-120$^{\circ}$ structure in the $ab$ plane~\cite{Hagihala2018}.
The 120$^{\circ}$ structure propagates antiferromagnetically along the $a$ and $c$ axes with a magnetic propagation vector ${\bm k}_{\rm mag}=(1/2,0,1/2)$.
Discussion of ground states in the kagome-triangular lattice model suggested that the identified 120$^{\circ}$ structure originates from a ferromagnetic intertube coupling, the Dzyaloshinskii-Moriya interaction, and a strong in-plane single-ion anisotropy.
In addition, it was found that the ground state of CsCrF$_{4}$ is close to the boundary on the magnetic phase diagram in the kagome-triangular lattice model~\cite{Hagihala2018}.
This suggests that small perturbations may induce various types of magnetic states in CsCrF$_{4}$.

Chemical substitution controls the magnetic state in CsCrF$_{4}$~\cite{Miura2011,Manaka2019}.
Thermodynamic and magnetic measurements for chemically substituted CsCr$_{1-x}$Fe$_{x}$F$_{4}$ and CsCr$_{1-x}$Al$_{x}$F$_{4}$ showed that the magnetic state is significantly influenced by the chemical composition~\cite{Manaka2019}.
An antiferromagnetic transition was clearly observed at 4.5~K for $x=0.06$ in the magnetic Fe-substituted compound, and the substituted superexchange bond Cr$^{3+}$-F$^{-}$-Fe$^{3+}$ enhanced the in-plane magnetic anisotropy. 
A glasslike transition appeared at about 5~K for the nonmagnetic Al-substituted compound.
In the present paper, we investigate long-range magnetic ordering in chemically substituted frustrated antiferromagnet CsCrF$_{4}$.
Magnetic structures of CsCr$_{0.94}$Fe$_{0.06}$F$_{4}$ and CsCr$_{0.98}$Al$_{0.02}$F$_{4}$ are identified by a combination of neutron powder diffraction experiments and magnetic structure analysis.
The most notable result is that the Fe-substitution effectively turns the ferromagnetic intertube coupling antiferromagnetic.

\section{Experimental details}
Polycrystalline samples of CsCr$_{0.94}$Fe$_{0.06}$F$_{4}$ and CsCr$_{0.98}$Al$_{0.02}$F$_{4}$ were prepared by a solid-state reaction method~\cite{Manaka2009,Manaka2019}.
The powder samples were loaded in vanadium-made containers, which were in turn installed in a conventional liquid $^{4}$He cryostat. 
Neutron diffraction measurements were performed at the high resolution powder diffractometer ECHIDNA~\cite{Echidna2006,Echidna2018} installed at the OPAL research reactor operated by the Australian Nuclear Science and Technology Organisation (ANSTO). 
We chose a Ge(331) monochromator to obtain neutrons with a wavelength of 2.4395~{\AA}, and used the open-open-$5'$ configuration.  
Temperatures were set at 10 K and 1.5 K.
The neutron diffraction data were analyzed by the Rietveld method with the FULLPROF software~\cite{PhysicaB192}. 
Candidates for magnetic structures compatible with the lattice symmetry were obtained by the SARA$h$ software~\cite{PhysicaB276}.
We used the VESTA software~\cite{vesta} for drawing the crystal structures and magnetic structures.

\section{Results and analysis}
\begin{figure}[tbp]
\includegraphics[scale=1]{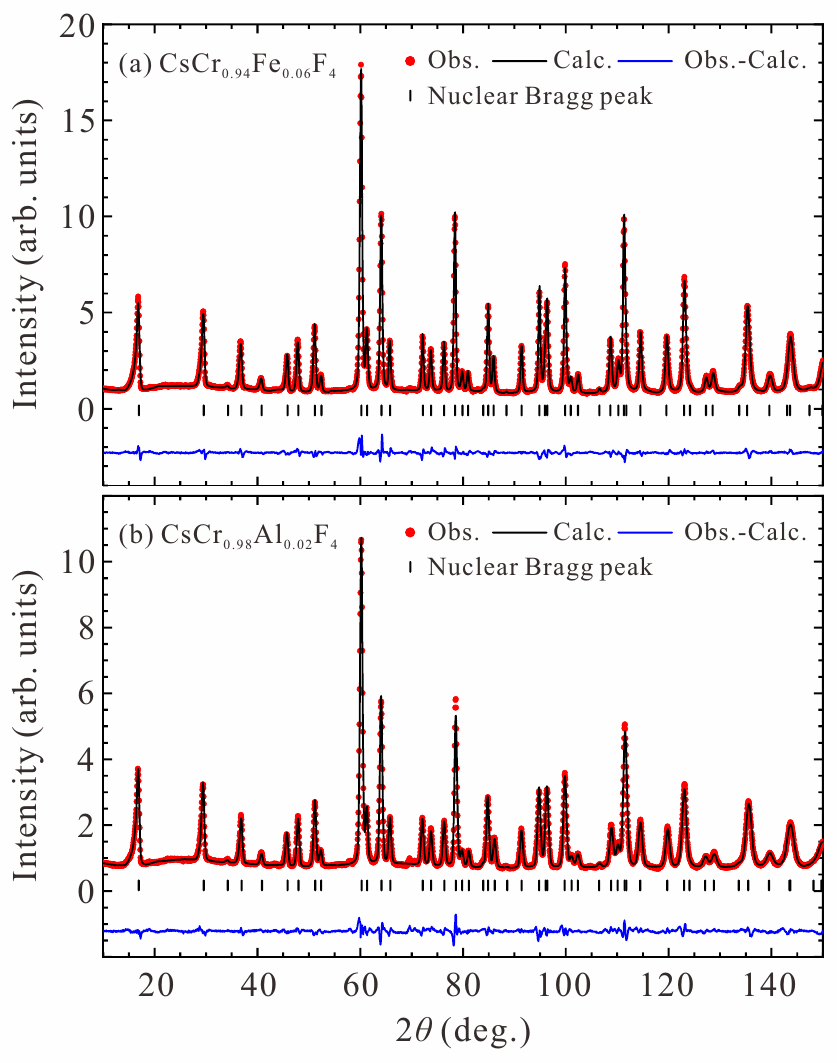}
\caption{Neutron powder diffraction profiles for 
(a) CsCr$_{0.94}$Fe$_{0.06}$F$_{4}$ and (b) CsCr$_{0.98}$Al$_{0.02}$F$_{4}$ measured at 10 K.
Red squares and black curves show the experimental data and simulations, respectively.
Vertical bars indicate the position of the nuclear Bragg peaks. Solid curves below the
bars show the difference between the data and simulations.}
\label{fig2}
\end{figure}

Figures~\ref{fig2}(a) and \ref{fig2}(b) show neutron powder diffraction profiles at 10~K for CsCr$_{0.94}$Fe$_{0.06}$F$_{4}$ 
and CsCr$_{0.98}$Al$_{0.02}$F$_{4}$, respectively.
These are reasonably  fitted by the hexagonal structure with the space group $P\overline{6}2m$.
Obtained profile factors are $R_{\rm wp}=7.33${\%} and $R_{\rm e}=3.70${\%} for CsCr$_{0.94}$Fe$_{0.06}$F$_{4}$, and $R_{\rm wp}=9.26${\%} and $R_{\rm e}=5.01${\%} for CsCr$_{0.98}$Al$_{0.02}$F$_{4}$.
The refined lattice and structural parameters (Table~\ref{tb:lattice_parameter}) are consistent with the previous results measured by x-ray powder diffraction experiments~\cite{Miura2011,Manaka2019}.
We conclude that the crystal structures are retained at low temperatures.

\begin{table}[htbp]
\caption{Results of the structural refinement in space group $P\overline{6}2m$ of the neutron powder diffraction profiles measured at $T=10$ K for CsCr$_{0.94}$Fe$_{0.06}$F$_{4}$ and 
CsCr$_{0.98}$Al$_{0.02}$F$_{4}$.}
\label{tb:lattice_parameter}
\begin{tabular}{cc}
\hline 
\hline
 & CsCr$_{0.94}$Fe$_{0.06}$F$_{4}$  \\ \hline
 $a$ (\AA) & 9.56402(7) \\ 
 $c$ (\AA) & 3.85832(3) \\ 
 Cs ($3g$) & (0.57202(12),  0,  1/2) \\
 Cr and Fe ($3f$) & (0.22456(23),  0,  0)  \\
 F ($3f$) & (0.83226(14),  0,  0) \\
 F ($3g$) & (0.22024(15),  0,  1/2)  \\
 F ($6j$) & (0.43914(10),  0.16142(11),  0)  \\
\hline
 &  CsCr$_{0.98}$Al$_{0.02}$F$_{4}$ \\ \hline
 $a$ (\AA) & 9.56886(6) \\ 
 $c$ (\AA) & 3.85059(3) \\ 
 Cs ($3g$) & (0.57203(17)  0,  1/2) \\
 Cr and Al ($3f$) & (0.22327(36),  0,  0) \\
 F ($3f$) & (0.83208(19),  0,  0) \\
 F ($3g$) & (0.22089(21),  0,  1/2) \\
 F ($6j$) & (0.43801(13),  0.16032(15),  0) \\
\hline \hline 
\end{tabular}\\
\end{table}

\begin{figure}[tbp]
\includegraphics[scale=1]{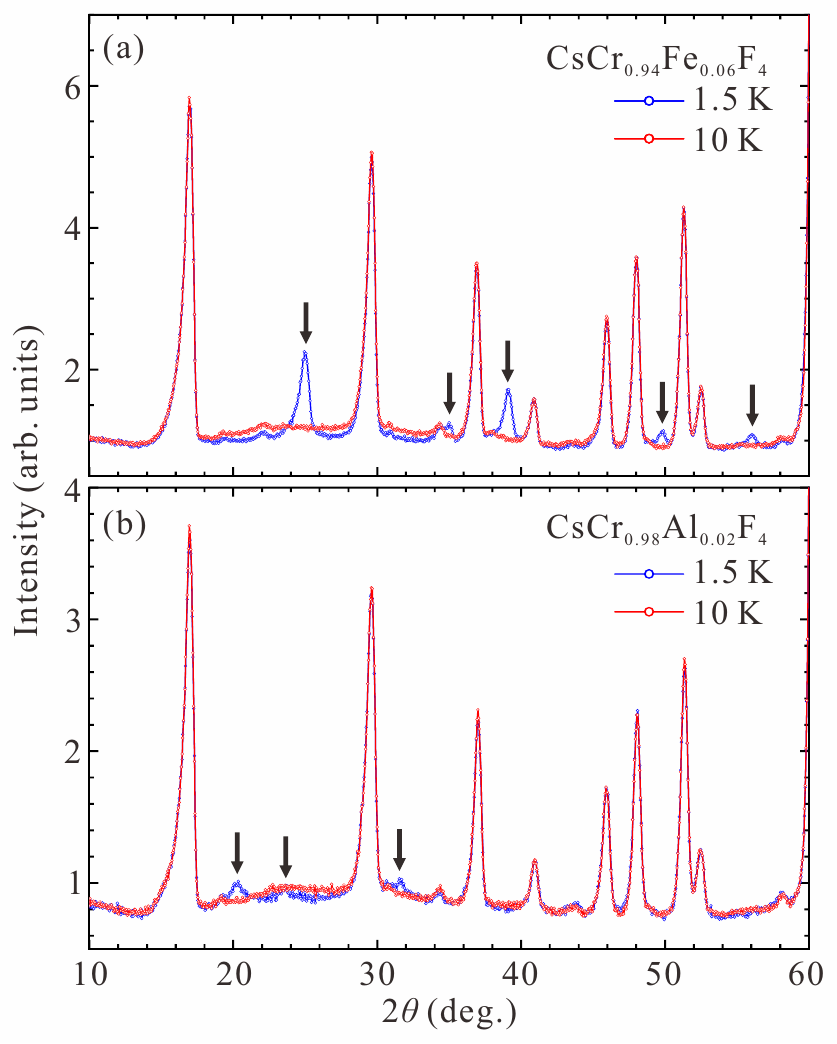}
\caption{Neutron powder diffraction profiles for 
(a) CsCr$_{0.94}$Fe$_{0.06}$F$_{4}$ and (b) CsCr$_{0.98}$Al$_{0.02}$F$_{4}$.
Blue and red marks are data measured at 1.5 and 10~K, respectively.
Arrows indicate the magnetic Bragg peaks.}
\label{fig3}
\end{figure}

Figure~\ref{fig3} shows neutron diffraction profiles at 1.5 and 10~K, which are below and above the transition temperature observed in the magnetic susceptibility data~\cite{Manaka2019}.
In both samples, diffuse scattering is observed at $10$~K in the range of $20^{\circ}\leq 2\theta \leq 40^{\circ}$, and it is suppressed below the transition temperature.
This behavior is the same as in CsCrF$_{4}$~\cite{Hagihala2018} and indicates that the short-range spin correlations develop at 10~K.
For CsCr$_{0.94}$Fe$_{0.06}$F$_{4}$, we clearly see additional peaks at 1.5~K, indicating the long-range magnetic order.
Additional peaks are also visible in CsCr$_{0.98}$Al$_{0.02}$F$_{4}$ even though their intensities are weak.
This result suggests that the anomaly observed at 5~K in the magnetic susceptibility measurements~\cite{Manaka2019} corresponds to weak long-range magnetic ordering rather than the glasslike transition.
Remarkably, the peaks for CsCr$_{0.94}$Fe$_{0.06}$F$_{4}$ are observed at different scattering angles from those in CsCrF$_{4}$, while the peak positions in CsCr$_{0.98}$Al$_{0.02}$F$_{4}$ are the same with CsCrF$_{4}$.
Indexing the observed magnetic Bragg peaks results in the magnetic propagation vectors ${\bm k}_{\rm mag}=(0,0,1/2)$ for CsCr$_{0.94}$Fe$_{0.06}$F$_{4}$ and ${\bm k}_{\rm mag}=(1/2,0,1/2)$ for CsCr$_{0.98}$Al$_{0.02}$F$_{4}$.

\begin{figure}[tbp]
\includegraphics[scale=1]{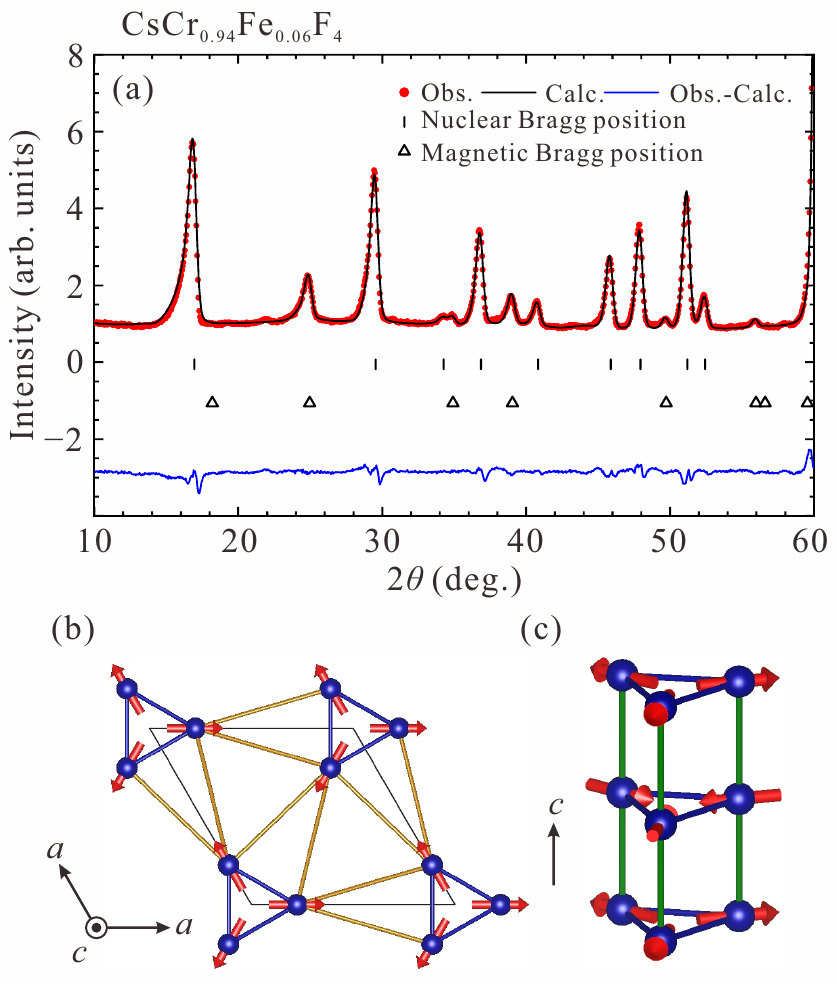}
\caption{(a) Neutron diffraction profile for 
CsCr$_{0.94}$Fe$_{0.06}$F$_{4}$ at 1.5 K.
Red squares and black curves show the experimental data and simulations, respectively.
Vertical bars and triangles indicate the position of the nuclear and magnetic Bragg peaks. 
Solid curves below the triangles show the difference between the data and simulations.
The magnetic structure of CsCr$_{0.94}$Fe$_{0.06}$F$_{4}$ 
with $\Gamma_{2}$ (b) in the $ab$ plane and (c) along the $c$ axis.}
\label{fig4}
\end{figure}

To determine magnetic structures that are compatible with the space group symmetry, we performed representation analysis. 
We assume that a magnetic structure is described by a single irreducible-representation (IR).
For CsCr$_{0.94}$Fe$_{0.06}$F$_{4}$, the representation analysis with the space group $P\overline{6}2m$ and the propagation vector ${\bm k_{\rm mag}}=(0,0,1/2)$ leads to five IRs. 
The details of the IRs and corresponding basis vectors are listed in Table~\ref{tb:IRFe}.
From the Rietveld refinement, a magnetic structure described by $\Gamma_{2}$ gives excellent agreement with the experimental data, as shown in Fig.~\ref{fig4}(a).
$R$ factors for the whole profile are $R_{\rm wp}=8.67${\%} and $R_{\rm e}=3.86${\%}.
The magnetic $R$ factor is $R_{\rm mag}=7.40${\%}.
Note that small peaks at $2\theta=22^{\circ}$ and 58$^{\circ}$ are likely due to nonmagnetic impurities because they are also observed at 10~K.
In the identified magnetic structure, the magnetic moments form a 120$^{\circ}$ structure in the $ab$ plane, and they propagate antiferromagnetically along the $c$ axis, as displayed in Figs.~\ref{fig4}(b) and \ref{fig4}(c).
The averaged magnitude of the magnetic moment is evaluated to be 1.66(1)~$\mu_{{\rm B}}$.
Since an effective magnetic moment for Cr$_{0.94}^{3+}$Fe$_{0.06}^{3+}$ is provided by $3.12~\mu_{\rm B}(=0.94\times3~\mu_{\rm B}+0.06\times5~\mu_{\rm B})$, the refined moment size is much smaller than the effective one.
This is the same result as that in the parent CsCrF$_{4}$~\cite{Hagihala2018}.
This implies that the magnetic moment strongly fluctuates even at 1.5~K due to the geometrical frustration and low dimensionality.

\begin{figure}[tbp]
\includegraphics[scale=1]{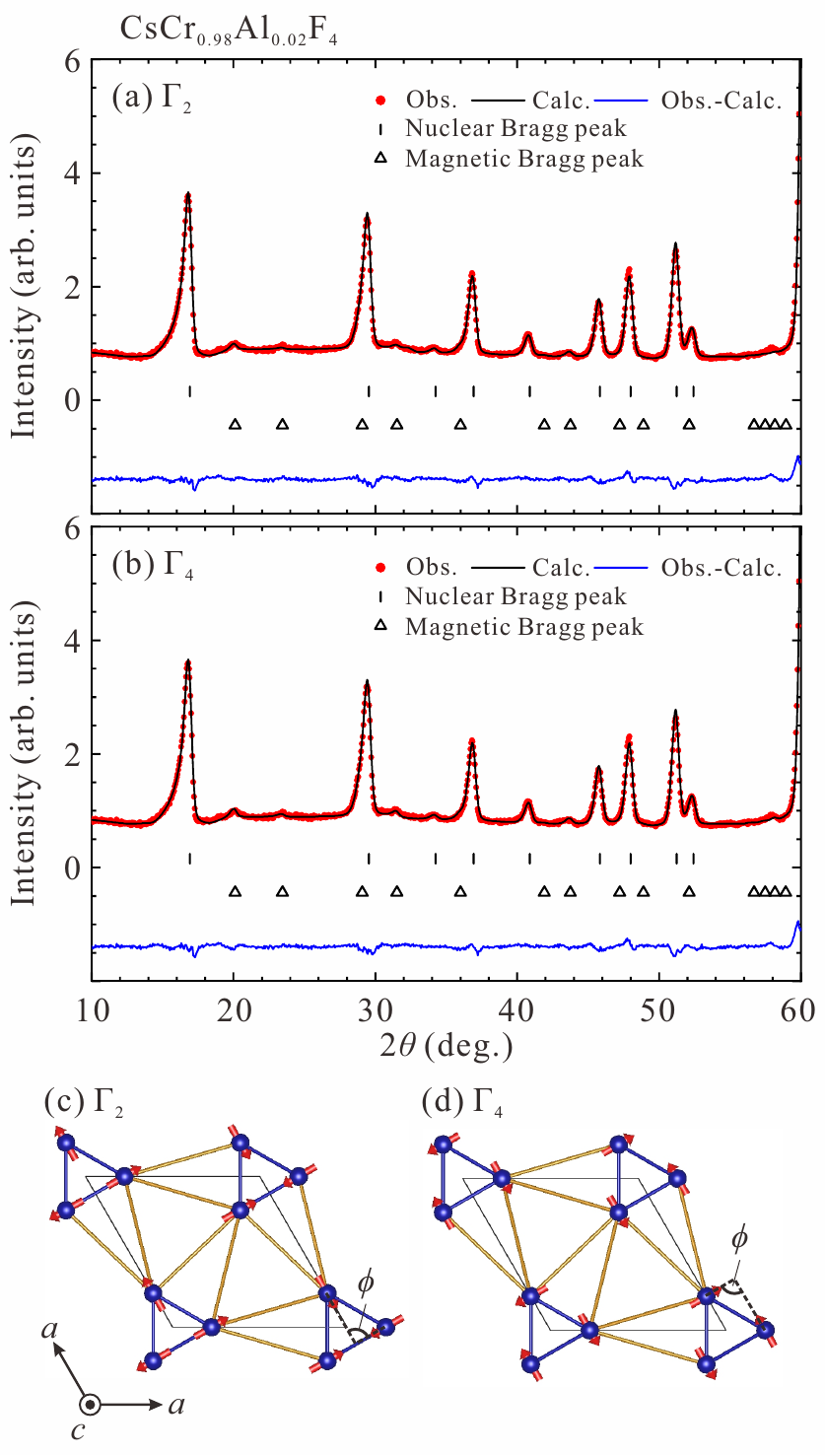}
\caption{Refined diffraction profiles for CsCr$_{0.98}$Al$_{0.02}$F$_{4}$ at 1.5 K.
The IRs of the magnetic structure for the simulations are (a) $\Gamma_{2}$ and (b) $\Gamma_{4}$, respectively.
Red squares and black curves show the experimental data and simulations, respectively.
Vertical bars and triangles indicate the position of the nuclear and magnetic Bragg peaks. 
Solid curves below the triangles show the difference between the data and simulations.
The magnetic structures of CsCr$_{0.98}$Al$_{0.02}$F$_{4}$ 
with (c) $\Gamma_{2}$ and (d) $\Gamma_{4}$.}
\label{fig5}
\end{figure}

For CsCr$_{0.98}$Al$_{0.02}$F$_{4}$, the representation analysis with the space group $P\overline{6}2m$ and the propagation vector ${\bm k}_{\rm mag}=(1/2,0,1/2)$ leads to splitting of the three equivalent Cr sites into the two nonequivalent Cr sites; site-1 $(0.2233, 0, 0)$, $(0.7767,0.7767,0)$, and site-2 $(0,0.2233,0)$.
Four and three IRs are associated with the site-1 and site-2, respectively.
The details of the IRs and corresponding basis vectors are listed in Table~\ref{tb:IRAl}.
From the Rietveld refinement, magnetic structures in $\Gamma_{2}$ and $\Gamma_{4}$ give satisfactory agreement with the experimental data, as shown in Figs.~\ref{fig5}(a) and \ref{fig5}(b).
$R$ factors for the whole profile are $R_{\rm wp}=9.59${\%} and $R_{\rm e}=5.05${\%} 
for $\Gamma_{2}$, and $R_{\rm wp}=9.65${\%} and $R_{\rm e}=5.06${\%} for $\Gamma_{4}$.
Magnetic $R$ factors are $R_{\rm mag}=18.8${\%} for $\Gamma_{2}$ and $R_{\rm mag}=15.8${\%} for $\Gamma_{4}$.
Note that it is hard to judge the optimal structure from these results because of the weak intensities of the magnetic Bragg peaks.
The identified magnetic structures with ${\bm k}_{\rm mag}=(1/2,0,1/2)$ are similar to that in the parent CsCrF$_{4}$, [see Figs.~\ref{fig5}(c) and \ref{fig5}(d)].
A quasi-120$^{\circ}$ structure is formed in the $ab$ plane, and it propagates antiferromagnetically along the $a$ and $c$ axes.
Relative angles $\phi$ between the moments at the sites 1 and 2, as indicated in Figs.~\ref{fig5}(c) and \ref{fig5}(d), are 92.4$^{\circ}$ for $\Gamma_{2}$ and 90.4$^{\circ}$ for $\Gamma_{4}$.
These angles deviate more from 120$^{\circ}$ than $119.5^{\circ}~(\Gamma_{2})$ and $108^{\circ}~(\Gamma_{4})$ in CsCrF$_{4}$~\cite{Hagihala2018}.
Refined magnitude of the magnetic moments is 1.04(7)$\mu_{{\rm B}}$ for $\Gamma_{2}$ and 1.14(6)$\mu_{{\rm B}}$ for $\Gamma_{4}$, 
As well as that of Cr$_{0.94}^{3+}$Fe$_{0.06}^{3+}$, the refined moment size of CsCr$_{0.98}$Al$_{0.02}$F$_{4}$ is much smaller than the effective moment $2.94~\mu_{\rm B}=(0.98\times 3~\mu_{\rm B})$ for Cr$_{0.98}^{3+}$Al$_{0.02}^{3+}$.
Thus, the ordered moments are strongly suppressed by the geometrical frustration and low dimensionality in the chemically substituted CsCrF$_{4}$.

\section{Discussion}
\begin{figure}[tbp]
\includegraphics[scale=1]{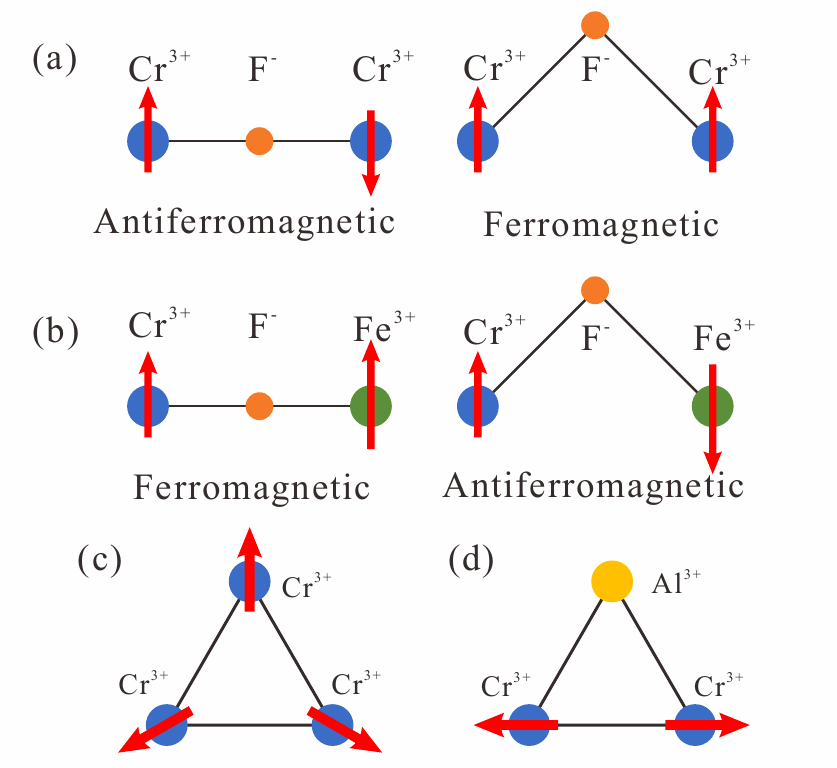}
\caption{(a) Superexchange interactions between the Cr$^{3+}$ ions in
180$^{\circ}$ and 90$^{\circ}$ bonds via the F$^{-}$ ion.
(b) Superexchange interactions between the Cr$^{3+}$ and Fe$^{3+}$ ions in
180$^{\circ}$ and 90$^{\circ}$ bonds via the F$^{-}$ ion.
[(c) and (d)] Spin structures with an antiferromagnetic interaction on an equilateral triangle.}
\label{fig6}
\end{figure}

In the magnetic structure analysis, the 120$^{\circ}$ structure having ${\bm k}_{\rm mag}=(0,0,1/2)$ is found for CsCr$_{0.94}$Fe$_{0.06}$F$_{4}$.
In the identified structure, the intertube spin configuration is different from that in the parent CsCrF$_{4}$, but the intratube structure is the same.
Let us discuss Fe-substitution effect on exchange interactions.
According to the Goodenough-Kanamori rules~\cite{Goodenough1955,Kanamori1959}, superexchange interactions between the Cr$^{3+}$ ions via the F$^{-}$ ion are antiferromagnetic for a 180$^{\circ}$ bond and ferromagnetic for a 90$^{\circ}$ bond as displayed in Fig.~\ref{fig6}(a).
Once the Cr$^{3+}$ ion is substituted by the Fe$^{3+}$ ion, the superexchange interactions in the 180$^{\circ}$ and 90$^{\circ}$ bonds are turned into ferromagnetic and antiferromagnetic ones, respectively [Fig.~\ref{fig6}(b)].
Since the bond angles of the nearest-neighbor exchange paths along the $c$ axis, $J_{0}$, and in the $ab$ plane, $J_{1}$, (see Fig.~\ref{fig1}) are 178$^{\circ}$ and 148$^{\circ}$ in CsCrF$_{4}$~\cite{Hagihala2018}, their exchange interactions likely turn antiferromagnetic into ferromagnetic by the Fe-substitution.
However, the identified magnetic structure in the triangular tube retains the same structure as that in CsCrF$_{4}$.
This means that the bond substitution in the intratube coupling has no real effect other than to create a small number of ferromagnetically coupled pairs.

On the contrary, the spin configuration between the spin tubes is totally different from that in CsCrF$_{4}$.
In CsCr$_{0.94}$Fe$_{0.06}$F$_{4}$, the magnetic propagation vector in the kagome-triangular plane ${\bm k}_{\rm 2D}$ is $(0, 0)$. 
It contrasts with ${\bm k}_{\rm 2D}=(1/2,0)$ in CsCrF$_{4}$.
This indicates that the substitution drastically changes the ground state even though intertube exchange paths are complicated.
In fact, according to the phase diagram of magnetic structures for the kagome-triangular lattice model~\cite{Hagihala2018,Seki2015}, 120$^{\circ}$ structures having ${\bm k}_{\rm 2D}=(0,0)$ and $(1/2,0)$ require antiferromagnetic and ferromagnetic intertube couplings, respectively.
We note that in the phase diagram the variation of the in-plane anisotropy described in Ref.~\cite{Manaka2019} does not change the intertube spin configuration.
Therefore, we conclude that in the magnetic Fe substitution the ground state is modified due to the evolution of the intertube coupling $J_{2}$ from ferromagnetic to antiferromagnetic.

The magnetic structure in CsCr$_{0.98}$Al$_{0.02}$F$_{4}$ is not changed drastically, even though the relative angle $\phi$ between the spins deviates more from 120$^{\circ}$ than that in the parent CsCrF$_{4}$.
On the basis of the classical vector-spin model, spins with the antiferromagnetic interaction form a 120$^{\circ}$ structure on an equilateral triangle, as displayed in Fig.~\ref{fig6}(c).
Substituting the Cr$^{3+}$ ion by the Al$^{3+}$ ion creates a spin vacancy in the triangle.
This likely induces the remaining two spins to align antiferromagnetically [Fig.~\ref{fig6}(d)].
Consequently, the Al substitution breaks a 120$^{\circ}$ structure locally.
However, the spin vacancy only produces a small effect on the ground state of CsCrF$_{4}$, and therefore the quasi-120$^{\circ}$ structure is still realized globally in CsCr$_{0.98}$Al$_{0.02}$F$_{4}$.

\section{Conclusion}
In conclusion, we have studied magnetic orders in magnetic Fe- and nonmagnetic Al-substituted CsCrF$_{4}$ through a neutron powder diffraction experiment.
Magnetic structure analysis reveals that the Fe-substituted sample exhibits a 120$^{\circ}$ 
structure having ${\bm k}_{\rm mag}=(0,0,1/2)$, and the Al-substituted one has a 
quasi-120$^{\circ}$ structure having ${\bm k}_{\rm mag}=(1/2,0,1/2)$.
Importantly, the magnetic structure in CsCr$_{0.94}$Fe$_{0.06}$F$_{4}$ differs from that in the parent CsCrF$_{4}$.
This result suggests that the ground state in CsCrF$_{4}$ is more sensitive to magnetic rather than nonmagnetic substitution on the Cr site.
Further studies of magnetic excitation for the Fe-substituted CsCrF$_{4}$ would be important to elucidate the spin interactions.

\section*{Acknowledgments}
We acknowledge the support of the Australian Centre for Neutron Scattering, Australian 
Nuclear Science and Technology Organisation, in providing the neutron research facilities 
used in this work.
The neutron diffraction experiments performed using ECHIDNA at ANSTO, Australia were supported by the General User Program for Neutron Scattering Experiments, Institute for Solid State Physics, The University of Tokyo (Proposals No. 18520), at JRR-3, Japan Atomic Energy Agency, Tokai, Japan.
S.H. was supported by the Japan Society for the Promotion of  Science through the Leading Graduate Schools (MERIT).

\begin{table}[htbp]
\caption{Basis vectors for the space group $P\overline{6}2m$ with ${\bm k}_{\rm mag} = (0,0,1/2)$. 
The atoms are defined according to Cr1: $(0.2246,0,0)$, 
Cr2: $(0,0.2246,0)$, Cr3: $(0.7754,0.7754,0)$.}
\centering
\begin{tabular}{cclll}
\hline \hline
IRs & & \multicolumn{3}{c}{Basis Vectors [$m_{a}$ $m_{b}$ $m_{c}$]} \\ 
 & & Cr1 & Cr2 & Cr3 \\ \hline \hline
$\Gamma_{2}$ & $\Psi_{1}$ & [1~0~0] & [0~1~0] & [-1~-1~0] \\
 & & & & \\
$\Gamma_{3}$ & $\Psi_{2}$ & [0~0~1] & [0~0~1] & [0~0~1] \\
 & & & & \\
$\Gamma_{4}$ & $\Psi_{3}$ & [1~2~0] & [-2~-1~0] & [1~-1~0] \\
 & & & & \\
$\Gamma_{5}$ & $\Psi_{4}$ & [0~0~2] & [0~0~-1] & [0~0~-1] \\
& $\Psi_{5}$ & [0~0~0] & [0~0~-$\sqrt{3}$] & [0~0~$\sqrt{3}$] \\
 & & & & \\
$\Gamma_{6}$ & $\Psi_{6}$ & [2~0~0] & [0~-1~0] & [1~1~0] \\
 & $\Psi_{7}$ & [0~1~0] & [$1/2$~$1/2$~0] & [-$1/2$~0~0] \\
 & & & +$i$[$\sqrt{3}/2$~$\sqrt{3}/2$~0] &  +$i$[$\sqrt{3}/2$~0~0] \\
 & $\Psi_{8}$ & [0~0~0] & [0~-$\sqrt{3}$~0] & [-$\sqrt{3}$~-$\sqrt{3}$~0] \\
 & $\Psi_{9}$ & [$1/2$~$1/2$~0] & [-$1/2$~0~0] & [0~1~0] \\
  & & +$i$[-$\sqrt{3}/2$~-$\sqrt{3}/2$~0] & +$i$[-$\sqrt{3}/2$~0~0] & \\
\hline \hline  
\end{tabular}
\label{tb:IRFe}
\end{table}

\begin{table}[htbp]
\caption{Basis vectors for the space group $P\overline{6}2m$ with 
${\bm k}_{\rm mag}=( 1/2,0,1/2)$. 
The atoms are defined according to Cr1: $(0.2233,0,0)$, 
Cr2: $(0,0.2233,0)$, Cr3: $(0.7767,0.7767,0)$.}
\centering
\begin{tabular}{ccll}
\hline \hline
IRs & & \multicolumn{2}{c}{Basis Vectors [$m_{a}$ $m_{b}$ $m_{c}$]} \\ \hline\hline
 & & Cr1 & Cr2  \\ \hline
$\Gamma_{1}$ & $\Psi_{1}^{(1)}$ & [0~0~1] & [0~0~1] \\
 & & & \\
$\Gamma_{2}$ & $\Psi_{2}^{(1)}$ & [1~0~0] & [1~1~0] \\
                     & $\Psi_{3}^{(1)}$ & [0~1~0] & [0~-1~0] \\
 & & & \\
$\Gamma_{3}$ & $\Psi_{4}^{(1)}$ & [0~0~1] & [0~0~-1] \\
 & & & \\
$\Gamma_{4}$ & $\Psi_{5}^{(1)}$ & [1~0~0] & [-1~-1~0] \\
                     & $\Psi_{6}^{(1)}$ & [0~1~0] & [0~1~0] \\ \hline
 & & Cr3 &   \\ \hline
$\Gamma_{2}$ & $\Psi_{1}^{(2)}$ & [0~-1~0] &  \\
 & & & \\
$\Gamma_{3}$ & $\Psi_{2}^{(2)}$ & [0~0~2] &  \\
 & & & \\
$\Gamma_{4}$ & $\Psi_{3}^{(2)}$ & [2~1~0] & \\
\hline \hline
\end{tabular}
\label{tb:IRAl}
\end{table}

\end{document}